\documentstyle[11pt]{article}
\newcommand{\beq}{\begin{equation}}
\newcommand{\eeq}{\end{equation}} 

\newcommand{\beqa}{\begin{eqnarray}}
\newcommand{\eeqa}{\end{eqnarray}}

\def\half{\frac{1}{2}}
\def\opone{\leavevmode\hbox{\small1\kern-3.8pt\normalsize1}}
\def\ua{\uparrow}  
\def\da{\downarrow}  
\def\B{{\cal B}}
 
\begin{document}

\title{Bell inequality, Bell states and maximally entangled states for n qubits}
\author
{N. Gisin and H. Bechmann-Pasquinucci\\
\protect\small\em Group of Applied Physics, University of Geneva, 
1211 Geneva 4, Switzerland}
\date{\today}

\maketitle

\begin{abstract}
First, we present a Bell type inequality for n qubits, assuming that m out of the n qubits 
are independent. Quantum mechanics violates this inequality
by a ratio that increases exponentially with m. Hence an experiment on n qubits violating 
of this inequality sets a lower bound on the number m of 
entangled qubits. Next, we propose a definition of maximally
entangled states of n qubits. For this purpose
we study 5 different criteria. Four of these criteria are found compatible. For any number 
n of qubits, they determine an orthogonal basis consisting of maximally entangled states
generalizing the Bell states.
\end{abstract}

\section{Introduction}\label{introduction}
We consider n qubits, all spatially separated from each other. The purpose of this
paper is to find characterizations of maximally entangled states of the n qubits. We
assume that the n qubits are equivalent (there is no privileged qubit(s)), hence we
restrict ourself to symmetric states. We 
test the following 5 criteria that could, a priori, characterize such Symmetric Maximally
Entangled (SME) states:
\begin{enumerate}
\item maximally entangled states violate Bell inequality \cite{Bell1964} maximally,
\item maximally entangled states are maximally fragile,
\item whenever m<n qubits are measured (in some appropriate basis), the outcomes determine a maximally entangled state
of the $m-n$ remaining qubits,
\item the mutual information of measurement outcomes is maximal,
\item all partial traces of maximally entangled states are maximally mixed.
\end{enumerate}
The Bell inequality we use is presented in the next section and the above 
criteria are made more precise and analysed in section 3.
The result of our analysis is that the last one of the above criteria is not a good
characterization. However, we find that all the other 4 properties are compatible and
define SME states. By local transformation, the SME states generate a basis of the entire
Hilbert space.

\section{Bell-Klyshko inequality for n qubits}\label{sectBell}
In this section we briefly present a generalization to n qubits of the 
Bell-CHSH \cite{CHSH} inequality well known in the 2 qubits case. This generalization 
was first presented by N.D. Klyshko and A.V. Belinskii \cite{Klyshko}.

Let us start with a brief review of the well known 2 qubits case. Let $a=\pm1$ and $a'=\pm1$
denote 2 possible outcomes of 2 possible measurements on the first qubit and similarly 
$b=\pm1$ and $b'=\pm1$ for the second qubit. Let us consider the following
linear combination of joint results:
\beq\label{F2}
F_2\equiv ab+ab'+a'b-a'b'=(a+a')b + (a-a')b' \le 2
\eeq
The above inequality holds because either $a=a'$ and the second term vanishes or $a=-a'$ and the first term
vanishes. In the first case the inequality follows because $|a+a'|\le2$ and $|b|\le1$,
and similarly in the second case. Assuming "local realism", the a's and b's are
independent and the probabilities of joint results are simply means of products.
Using the linearity of the mean operator $M$ and
denoting $E_{indep.}(a,b)=M(a.b)$ the expectation value of the products of the results of the
experiments $a$ and $b$, one obtains the traditionally Bell-CHSH inequality \cite{CHSH}:
\beq\label{CHSHB2}
E_{indep.}(a,b)+E_{indep.}(a,b')+E_{indep.}(a',b)-E_{indep.}(a',b')\le2
\eeq
The quantum mechanical description associates to the first measurement the Pauli 
matrix $\vec a\vec\sigma$ with normalized 3-dim vectors $\vec a$, and similarly
for the other measurements. The mean correlation in state $\psi$ is given by the 
quantum expectations: $E_{QM}(a,b)=\langle\vec a\vec\sigma\otimes\vec b\vec\sigma\rangle_\psi$. Hence, the
Bell-CHSH inequality involves the Bell operator:
\beq
\B_2=\vec a\vec\sigma\otimes\vec b\vec\sigma+\vec a\vec\sigma\otimes\vec b'\vec\sigma+
\vec a'\vec\sigma\otimes\vec b\vec\sigma-\vec a'\vec\sigma\otimes\vec b'\vec\sigma
\eeq
A straightforward computation shows that $\B_2^2\le8$. Accordingly, the largest eigenvalue
of $\B_2$ is $2\sqrt{2}$ and the Bell-CHSH inequality can be violated by quantum state
by a maximal factor of $\sqrt{2}$ \cite{Cirelson}.

The above brief presentation of the standard Bell-CHSH inequality, valid for 2 qubits,
motivates the following generalization for n qubits which is defined recursively.
Let $a_n=\pm1$ and $a_n'=\pm1$ denote 2 possible outcomes of 2 possible measurements on 
the nth qubit and define:
\beq\label{CHSHBn}
F_n\equiv\half(a_n+a_n')F_{n-1} + \half(a_n-a_n')F_{n-1}'\le2
\eeq
where $F_n'$ denote the same expression $F_n$ but with all the $a_j$ and $a_j'$ 
exchanged. The above inequality (\ref{CHSHBn}) holds for precisely the same reason
as the standard 2 qubit Bell-CHSH inequality, as presented below eq. (\ref{F2}). 
Using again the linearity of the mean, one obtains the Bell-Klyshko inequality for n qubits:
\beq \label{inequality}
E_{indep.}(F_n(a_1,...,a_n))\le 2
\eeq
Similarly to the 2 qubit case, we define the Bell operator for n qubits:
\beq
\B_n=\B_{n-1}\otimes\half(\vec a_n\vec\sigma+\vec a_n'\vec\sigma)
+\B_{n-1}'\otimes\half(\vec a_n\vec\sigma-\vec a_n'\vec\sigma)
\eeq
An upper bound on the eigenvalues of $\B_n$ is given by the following lema:
\beq
\B_n^2\le2^{n+1}
\label{Bn}
\eeq
Again, the proof follows closely the one for the 2 qubit case:
\beqa
\B_n^2&=&\B_{n-1}^2\otimes\half(1+\vec a_n\vec a_n')+(\B_{n-1}')^2\otimes\half(1-\vec a_n\vec a_n')
+[\B_{n-1},\B_{n-1}']\otimes\half i (\vec a_n\wedge\vec a_n')\vec\sigma \\ \nonumber
&\le&(1+|\vec a_n\wedge\vec a_n'|)2^n \\ \nonumber
&\le& 2^{n+1}
\eeqa
The maximally entangled states discussed in the next section saturate the above bound. Hence,
the largest eigenvalue of $\B_n$ is $2^{(n+1)/2}$ and the Bell-Klyshko inequality for n
qubits can be violated by quantum states by a maximal factor of $2^{(n-1)/2}$.

The maximal violation of the inequality (\ref{inequality}) is larger than the one presented by Mermin 
\cite{Mermin1990} for
even numbers n of qubits and is larger than the inequality derived by Ardehali 
\cite{Ardehali} for
odd numbers of qubits.

The inequality (\ref{inequality}) is symmetric among the n qubits. For example, for 3 qubit it
reads: $E(a,b,c')+E(a,b',c)+E(a',b,c)-E(a',b',c') \le 2$.

An interesting property of the inequality (\ref{inequality}) is that if one assumes that
only $m<n$ qubits have independent elements of reality, then, combining the inequalities
(\ref{CHSHBn}) and (\ref{Bn}), one obtains:
\beq
E_{m~indep.~qubits}(F_{n})\le2^{(n-m+1)/2}
\eeq
Indeed, applying recursively the definition (\ref{CHSHBn}) one obtains (see Appendix B):
\beq
F_n = \frac{1}{4}(F_{n-m}+F_{n-m}')F_m + \frac{1}{4}(F_{n-m}-F_{n-m}')F_m'
\label{Fnm}
\eeq
Hence, if the $m$ last qubits are independent, then $F_m\le2$ and $F_m'\le2$,
thus $F_n\le F_{n-m}$. Finally, if the $n-m$ first qubits are maximally entangled, then
$E_{m~indep.~qubits}(F_{n})=E_{QM}(F_{n-m})\le2^{(n-m+1)/2}$. 
Note that by symmetry the same result hold if any $m$ qubits are independent.
Accordingly, from an experimental measurement result of $E(F_n)$ between $2^{(n-m+1)/2}$
and $2^{(n-m+2)/2}$ one can infer that at least $m$ qubits are entangled \cite{Svetlichny,Zukowsky}.
For example consider the following 3-qubit mixed state with only 2-qubit entanglement
\cite{Brassard}: $\rho=\half(P_S\otimes P_+ + P_+\otimes P_S)$ where $P_S$ denotes the 
singlet state of 2 qubits and $P_+=|+\rangle_{zz}\langle+|$ the "up"
state of a single qubit. Then the maximal
violation of the generalized Bell-CHSH inequality (\ref{inequality}) equals
$E_{QM}(F_3)=2(1+\sqrt{2})$. This value is reached for
$\alpha=-\alpha'=\gamma=-\gamma'=\pi/8$, $\beta=\pi$, $\beta'=\pi/2$ where $\alpha$ and
$\alpha'$ are the angles defining the directions $a$ and $a'$, respectively, in the $xz$
plane (angle 0 corresponding to the $z$ direction) and
similarly for the 4 other angles. From this value one concludes rightly that the
state $\rho$ contains 2-qubit entanglement.

\section{Maximally entangled states of n qubits}
In this section we consider several criteria to characterize Symmetric
Maximally Entangled (SME) states of n qubits.

\subsection{Maximal violation of Bell inequality}
Maximally entangled states should maximally violation Bell inequality. This criteria is
not precise, because there are infinitely many versions of Bell inequality. In this
subsection we consider the Klyshko version presented in section \ref{sectBell}, because it
is the natural generalization of the Bell-CHSH inequality which nicely characterizes
the 2 qubits Bell states.
The generalized Bell-Klyshko inequality (\ref{inequality}) for n qubits is maximally 
violated by the GHZ \cite{GHZ} state $|\ua...\ua\rangle+|\da...\da\rangle$ for $\vec a_j$ regularly 
distributed in the xy plane
(on the Poincar\'e sphere) with angles $(j-1)(-1)^{n+1}\frac{\pi}{2n}$ with respect
to the x-axis and $a_j'\perp a_j$. 
More generally, the state $\alpha|\ua...\ua\rangle+\beta|\da...\da\rangle$ with real
$\alpha$ and $\beta$ satisfying $\alpha^2+\beta^2=1$ violate the n qubit inequality 
(\ref{inequality}) by a factor $\alpha\beta2^{(n-1)/2}$.

\subsection{Maximal entanglement as maximal fragility}\label{secfragile}
It seems natural to assume that entanglement is fragile. More specifically, if the
qubits are subject to noise, acting independently on each of them, the more
entangled states are more affected than the less entangled ones. Noise acting on a
qubit can be modelled by a fluctuating Hamiltonian: $H=\vec\beta\vec\sigma$ 
where the 3 components of the vector $\vec\beta$ are independent Wiener processes
(ie white noise). 
Accordingly, the density matrix of the n qubits follows the master equation:
\beq
\dot\rho_t=\sum_{j=1}^n(\vec\sigma_j\rho_t\vec\sigma_j-3\rho_t)
\eeq
where the Pauli matrices $\vec\sigma_j$ act on the jth qubit. 
Now, a state $\psi$ is fragile if under
the above evolution it quickly drift away, that is if
$|\langle\psi|\dot P_\psi|\psi\rangle|=|\sum_{j=1}^n\langle\vec\sigma_j\rangle_\psi^2-3n|$ is large.
Accordingly, a state $\psi$ is maximally fragile iff $\langle\vec\sigma_j\rangle_\psi^2=0$ for all j.
This condition is equivalent to the 1-qubit partial state being equal to the maximally
mixed state: 
\beq \label{rho1}
\rho_1\equiv Tr_{2..n}(P_\psi)=\opone/2
\eeq

Another type of fragility, natural for entanglement, is fragile under measurement
interactions: if one or some qubits are measured (in the computational basis) on a
maximally entangled state, then all the entanglement is destroyed, ie the measurement
projects the state onto a product state.

The GHZ states $|0...0\rangle\pm|1...1\rangle$ satisfy both of these 2 criteria of maximal fragility.

\subsection{Distribution of entangled stated}\label{secdistribute}
Another criteria for maximal entanglement could be the following. A state of n qubits is
maximally entangled if any k<n holders of a qubit can distribute to the k-n other qubit holders
a maximally entangled state. Note that this criteria applies recursively to larger and
larger number of qubits, starting from the well known Bell states for 2 qubits. 

At first this criteria may seem in contradiction with the one of the previous
sub-section \ref{secfragile}. However, this is not the case, provided the measurement
is not done in the computational basis (the "$z$-basis"), but in the "$x$-basis". Indeed, 
using the formula of the appendix the GHZ states can be rewritten as:
\beqa
|0,n\rangle_z+|n,n\rangle_z&=&(|0,k\rangle_z+|k,k\rangle_z)\otimes(|0,n-k\rangle_z+|n-k,n-k\rangle_z) \\ \nonumber
&& + (|0,k\rangle_z-|k,k\rangle_z)\otimes(|0,n-k\rangle_z-|n-k,n-k\rangle_z) \\
&=&\sum_{j=0}^{k/2}|2j,k\rangle_x\otimes(|0,n-k\rangle_z+|n-k,n-k\rangle_z) \\ \nonumber
&&+ \sum_{j=0}^{(k+1)/2}|2j+1,k\rangle_x\otimes(|0,n-k\rangle_z-|n-k,n-k\rangle_z)
\eeqa
Hence, if k qubits are measured in the $x$-basis and an even number of 1 are obtained,
then the remaining n-k qubits are in the GHZ state $|0,n-k\rangle_z+|n-k,n-k\rangle_z$. Else, if an
odd number of 1 are obtained, the n-k qubits are in the other GHZ state
$|0,n-k\rangle_z-|n-k,n-k\rangle_z$.

\subsection{Mutual information of measurement outcomes on maximally entangled states}\label{secmutinfo}
Let $a_j$ denote the results of simultaneous measurements of all qubits 
in the computational basis. A natural
criterion for maximal entanglement is that the mutual information of the n random
variables $a_j$ is maximum. Recall that $I(\{a_j\})\equiv \sum_j H(a_j)-H(a_1,...a_n)$,
where $H$ is the entropy function $H(x)\equiv-M\left(\log(p(x))\right)$. Using the
symmetry among the $a_j$ and the chain rule \cite{Cover} $H(a_1,...,a_n)=H(a_1)+H(a_2,...,a_n|a_1)$,
one obtains:
\beq
I(\{a_j\})=(n-1)H(a_1)-H(a_2,...,a_n|a_1)
\eeq
Accordingly, the mutual
information $I(\{a_j\})$ is maximal if $H(a_1)$ is maximal and $H(a_2,...,a_n|a_1)$ is minimal.
Finally, $H(a_1)$ is maximal if both outcomes are equally probable, and 
$H(a_2,...,a_n|a_1)$ is minimal if the outcome $a_1$ fully determines all the results
$a_2$ to $a_n$. The GHZ states do clearly satisfy this criterion. Note that for 
$n\ge3$ they are the only symmetric states satisfying this criterion (for $n=2$, the third
symmetric Bell state does also satisfy this criterion). Note also the similarity between
this mathematical criteria and the more physical "maximally fragile" criteria of 
section \ref{secfragile}: maximizing $H(a_1)$ is equivalent to condition (\ref{rho1})
which represents maximal fragility under noise
and minimizing $H(a_2,...,a_n|a_1)$ is equivalent maximal fragility under 1-qubit
measurement in the computational basis.

\subsection{Partial states of maximally entangle states are maximally mixed?}\label{secpartial}
In subsection \ref{secfragile} we found that the 1-qubit partial state of a
maximally fragile states is the maximally mixed state. A more general criterion for
maximal entanglement could be that all partial states are maximally mixed. Since we 
assume all through this article symmetric states, maximally mixed states differ from
states represented by a multiple of the identity matrix: maximally mixed states are homogenous
mixtures of all the $n+1$ symmetric states. For example, the maximally mixed state of 
2-qubit is the mixture of the 3 symmetric Bell states, all 3 with the same weight.
Note that the density matrices representing $m$-qubits and $(n-m)$-qubits partial states of an
$n$-qubit state have the same spectra. Hence, only the partial states for 
$m=1,...,[\frac{n}{2}]$ can be maximally mixed (where $[\frac{n}{2}]$ is the 
largest integer smaller or equal to $\frac{n}{2}$).
Consequently, the m-qubit partial states of a SME state should,
according to this criteria, have exactly $m+1$ identical eigenvalues different from zero, for
all $m=1,...,[\frac{n}{2}]$. Note that the conditions for different m's are not independent.
Indeed, if the partial state for $m=[\frac{n}{2}]$ is maximally mixed, then all partial states
for smaller m's are also necessarily maximally mixed.

We found states of n-qubits satisfying this criterion for n=2,3,4 and 6. However,
no such state exists for n=5, nor for $n\ge7$! Hence, in general, no state satisfying this 
criterion exists. This is no surprise, since the number of constrains increases with $n$ 
much faster than the number of parameters defining symmetric states. 
This criterion is thus not suitable to characterize maximal entanglement. Indeed, if
entanglement can be measured, then maximally entangled states should exist for any number
of qubits.

For completeness we nevertheless list some examples of states satisfying this criterion:
\beqa
\psi_{3,\pm1}&=&|0,3\rangle\pm|3,3\rangle \\
\psi_{3,\pm2}&=&|0,3\rangle\pm|1,3\rangle-|2,3\rangle\mp|3,3\rangle \\
\psi_{4,\pm1}&=&-3|0,4\rangle\pm\sqrt{3}|1,4\rangle+|2,4\rangle\pm\sqrt{3}|3,4\rangle-3|4,4\rangle \\
\psi_{4,2}&=&|0,4\rangle+\frac{i}{\sqrt{3}}|2,4\rangle+|4,4\rangle \\
\psi_{6,\pm1}&=&|1,6\rangle\pm|5,6\rangle \\
\psi_{6,2}&=&-3|0,6\rangle+|2,6\rangle+|4,6\rangle-3|6,6\rangle \\
\psi_{6,\pm3}&=&\sqrt{2}|0,6\rangle\pm\frac{i}{2}|3,6\rangle+\sqrt{2}|6,6\rangle 
\eeqa
where $|j,n\rangle$
denotes the sum of all product states with $j$~ $|1\rangle$ and $n-j$~ $|0\rangle$ (see
Appendix A)
and where the first index of the $\psi$'s indicate the number of qubits and the second index
label the different states with maximally mixed partial states.

\section{Conclusion}\label{concl}
The Bell-Klyshko inequality for n qubits has been presented. Maximally
entangled quantum states violate this inequality by a factor that grows exponentially with n.
For even and odd numbers of qubits, this maximal violation is larger than for the inequality
devised by Mermin \cite{Mermin1990} and by Ardehali \cite{Ardehali}, respectively. If only $n-m$
qubits are assumed independent, then the same inequality leads to a higher bound. Hence, from
an experimental test of the Bell-Klyshko inequality for n qubits, one can infer a lower bound
on the number of entangled qubits.

Maximally entangled symmetric states of n qubits were analysed according to 5 different 
criteria. The criteria that the partial states should be maximally mixed is found to be
of limited value, since no such states exist neither for $n=5$ nor for $n\ge7$. However, all
4 other criteria are shown to be compatible. The conclusion is that the 2 GHZ states
\cite{GHZ} are the maximally entangled symmetric states: they violate Bell inequality maximally, they
are maximally fragile and maximize the mutual information. Moreover, 
depending on the measurement
bases, $m$ holders of qubits can either distribute maximally entangled qubits to their $n-m$
colleagues or leave them with product states.

Clearly all states obtained by local transformation of a maximally entangled state are
equally valid maximally states \cite{Popescu}, though not necessarily symmetric states. 
Starting from the GHZ state one thus obtains $2^n$ linearly independent maximally entangled
states. Hence, the maximally entangled states form an orthogonal basis for any number of
qubits, like the well known Bell states for 2 qubits. For example, in the 3 qubit case the
two GHZ states read $|1,1,1\rangle\pm|0,0,0\rangle$. By local transformation on the
first qubit the following pair of states obtains: $|0,1,1\rangle\pm|1,0,0\rangle$.
Acting similarly on the second and third qubits provides a total of 8 mutually
orthogonal states that form a basis of the 3-qubit Hilbert space.

In this letter we did not address the question of unicity. However, our experience leads us to
conjecture that the GHZ states and the states obtained from them by local transformations
are the unique states that violate maximally the Bell inequality (\ref{inequality}) and are the
unique states satisfying independently the criteria of sub-sections \ref{secfragile}, 
\ref{secdistribute} and \ref{secmutinfo}.
Hence, we conjecture that our 4 criteria are not only compatible, but equivalent
characterization of maximally entangled states.

\section*{Acknowledgments}
Stimulating discussions with A. Ekert, B. Huttner, S. Popescu, A. Zeilinger 
and M. \.Zukowski are acknowledged. In particular we thank M. \.Zukowski for bringing
references \cite{Klyshko} to our attention. H.B.-P. is supported by the Danish National Science 
Research Council (grant no. 9601645). This work profited also from support by the Swiss
National Science Foundation.

\section*{Appendix A: symmetric states of n qubits}
In this appendix we summarize some useful properties and notations for symmetric states. 
Let $|j,n\rangle$
denote the sum of all product states with $j$~ $|1\rangle$ and $n-j$~ $|0\rangle$. For example:
$|1,3\rangle\equiv|1,0,0\rangle+|0,1,0\rangle+|0,0,1\rangle$. Hence $|j,n\rangle$ is a $2^n$ component vector
representing a n-qubit state. The norms and inner products are given by:
\beq
\langle j,n|k,n\rangle=\delta_{j,k}C_j^n\equiv\delta_{j,k}\frac{n!}{(n-j)!j!}
\eeq
The decomposition of symmetric states of n qubits on symmetric states of m qubits states
is straightforward:
\beq
|j,n\rangle=\sum_{k=0}^j|k,m\rangle\otimes|j-k,m-n\rangle
\eeq
It is also relatively easy to change from one basis to another. For example, if
$|1\rangle_x\equiv|0\rangle_z+|1\rangle_z$ and $|0\rangle_x\equiv|0\rangle_z-|1\rangle_z$, then:
\beq
|j,n\rangle_z=\sum_{\ell=0}^n\big(\sum_{k=0}^{\frac{j-\ell}{2}}C_{j-2k}^\ell C_{2k}^{n-\ell}
- \sum_{k=0}^{\frac{j-\ell-1}{2}}C_{j-2k-1}^\ell C_{2k+1}^{n-\ell}\big) |\ell,n\rangle_x
\eeq
Accordingly the GHZ states expressed in the $x$-bases read:
\beq
|0,n\rangle_z+|n,n\rangle_z=\sum_{k=0}^{n/2}|2k,n\rangle_x
\eeq
\beq
|0,n\rangle_z-|n,n\rangle_z=\sum_{k=0}^{(n+1)/2}|2k+1,n\rangle_x
\eeq
Similarly, the GHZ states expressed in y-basis 
($|1\rangle_y\equiv|1\rangle_z+i|o\rangle_z$, $|0\rangle_y\equiv|0\rangle_z+i|1\rangle_z$) read:
\beq
|0,n\rangle_z\pm|n,n\rangle_z=\sum_{k=0}^n(i^k\pm i^{n-k})|k,n\rangle_y
\eeq

\section*{Appendix B: Proof of formula (\ref{Fnm})}
Let us prove it by induction on $m$. 
Defining $F_1(a)\equiv 2a$, formula  (\ref{Fnm}) holds for $m=1$ and,
recalling that $F_m''=F_m$, one has:
\beqa
F_n&=&\frac{F_m+F_m'}{4}F_{n-m} + \frac{F_m-F_m'}{4}F_{n-m}' \\
&=&\frac{F_m+F_m'}{4}\left(\frac{a_{m+1}+a_{m+1}'}{2}F_{n-(m+1)} + \frac{a_{m+1}-a_{m+1}'}{2}F_{n-(m+1)}'\right) \\ \nonumber
&+&\frac{F_m-F_m'}{4}\left(\frac{a_{m+1}'+a_{m+1}}{2}F_{n-(m+1)}' + \frac{a_{m+1}'-a_{m+1}}{2}F_{n-(m+1)}\right) \\ 
&=&\frac{F_m a_{m+1}'+F_m'a_{m+1}}{4}F_{n-(m+1)} + \frac{F_m a_{m+1}-F_m'a_{m+1}'}{4}F_{n-(m+1)}' \\
&=&\frac{F_{m+1}+F_{m+1}'}{4}F_{n-(m+1)} + \frac{F_{m+1}-F_{m+1}'}{4}F_{n-(m+1)}'
\eeqa
Hence, assuming (\ref{Fnm}) holds for $m$, it also holds for $m+1$.

\end{document}